# Negative thermal expansion in the Prussian Blue analog $Zn_3[Fe(CN)_6]_2$: X-ray diffraction and neutron vibrational studies

**S Adak**[1,2], **L L Daemen**[2] **and H Nakotte**[1,2]

[1]Department of Physics, New Mexico State University, Las Cruces, NM 88003, USA
[2]Los Alamos Neutron Science Center, Los Alamos National Laboratory, Los Alamos, NM 87545, USA

E-mail: sadak@lanl.gov

**Abstract**. The cubic Prussian Blue (PB) analog, $Zn_3[Fe(CN)_6]_2$, has been studied by X-ray powder diffraction and inelastic neutron scattering (INS). X-ray data collected at 300 and 84 K revealed negative thermal expansion (NTE) behaviour for this material. The NTE coefficient was found to be $-31.1 \times 10^{-6}$ $K^{-1}$. The neutron vibrational spectrum for $Zn_3[Fe(CN)_6]_2 \cdot xH_2O$, was studied in detail. The INS spectrum showed well-defined, well-separated bands corresponding to the stretching of and deformation modes of the Fe and Zn octahedra, all below 800 $cm^{-1}$.

## 1. Introduction
Negative thermal expansion (NTE), expansion on cooling or contraction upon heating, is an interesting phenomenon from both fundamental and applied physics viewpoints. It occurs in different classes of materials. In many cases, the physical mechanism behind this phenomenon is often not completely understood.

It seems to be "common knowledge" that many PB compounds display NTE, although quantitative information on thermal expansion coefficients is often missing. Furthermore, not all compounds in this family exhibit NTE. Therefore we started a comprehensive study of these compounds. The possibility of varying ion size and charge in this material offers an interesting playground for studying NTE systematically, and to investigate possible correlations with electronic and crystal structures. The linkage between the octahedral units in PB analogs with a linear cyanide ligand introduces more degrees of freedom in these cubic structures.

We studied the PB analog $Zn_3[Fe(CN)_6]_2 \cdot xH_2O$ in some detail. X-ray powder diffraction showed that this compound exhibits strong NTE behavior. We performed in depth studies of the inelastic neutron scattering (INS) spectrum of $Zn_3[Fe(CN)_6]_2 \cdot xD_2O$, and compared the experimental results with simple semi-empirical cluster calculations.

## 2. Experiment and Computation
A polycrystalline sample of $Zn_3[Fe(CN)_6]_2 \cdot xH_2O$ was prepared via standard chemical precipitation. All reagents were ACS quality and were used as-received without further purification. Inelastic neutron spectroscopic measurements were done using a deuterated sample, $Zn_3[Fe(CN)_6]_2 \cdot xD_2O$, prepared via the same synthesis route using $D_2O$ instead of $H_2O$.

X-ray diffraction (XRD) patterns at room temperature ($T_2$ = 300K) and low temperature ($T_1$ = 84K) were collected under vacuum for the powder sample on a Rigaku Ultima III X-ray diffractometer, operated in Bragg-Brentano geometry with monochromatic $CuK_\alpha$ ($\lambda$ = 1.54 Å) radiation. The lattice parameters were obtained from Rietveld refinements of the diffraction patterns using the General Structure Analysis System (GSAS) software [1].

The inelastic neutron scattering (INS) spectrum was collected using the Filter Difference Spectrometer (FDS) at Los Alamos Neutron Scattering Center (LANSCE). The interpretation of the INS spectrum was done with the help of simple cluster calculations (semi empirical, PM3 [2]) performed using HyperChem computational chemistry package [3], a-climax program [4], and the use of a few reference spectra [5].

### 3. Results & Discussions

$Zn_3[Fe(CN)_6]_2 \cdot xH_2O$ crystallizes into a cubic structure with space group $Fm\bar{3}m$. The structure consists of two types of octahedra ($Zn^{II}(NC)_6$ and $Fe^{III}(CN)_6$) arranged on a cubic lattice and linked by cyanide ligands. Water molecules, which contribute to the existence of structural disorder, appear in the space between the octahedrons. The structure is shown in figure 1.

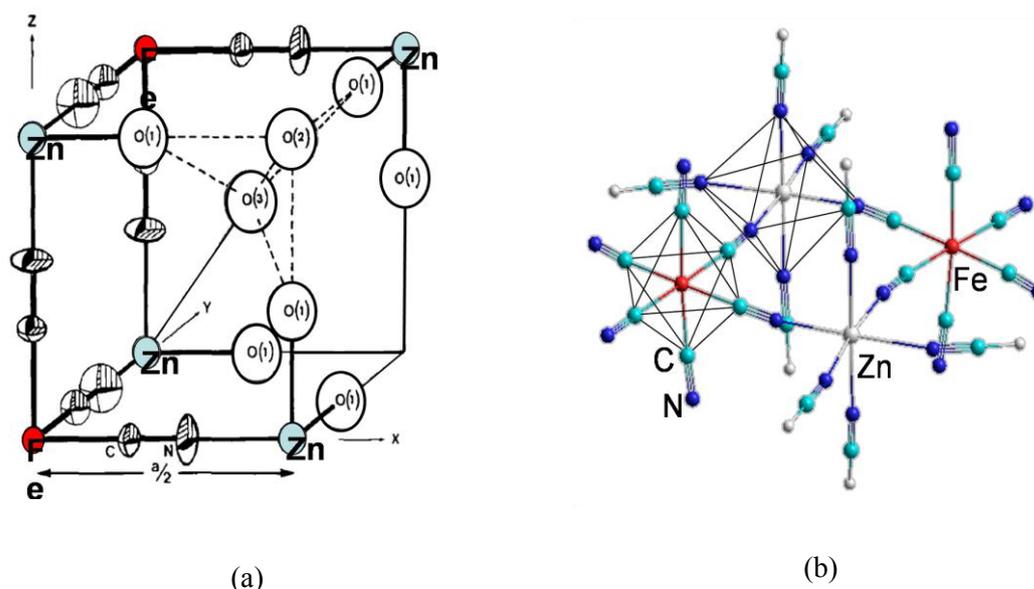

(a)        (b)

**Figure 1.** The structure of $Zn_3[Fe(CN)_6]_2$ showing (a) the possible sites for water in $Zn_3[Fe(CN)_6]_2 \cdot 14H_2O$ [6], and (b) a simple cluster model used to calculate the INS spectrum.

To determine if this PB analog shows NTE behavior, we studied the lattice parameters for the compound at two temperatures. We determined an average coefficient of thermal expansion (CTE) from the optimized lattice parameters obtained from the Rietveld refinement of X-ray diffraction (XRD) data collected at 300 and 84 K. For the refinements, crystallographic information available in the ICSD database [6] was used as a starting point. The lattice parameters, thermal parameters, atom positions, peak profiles, and background parameters were refined. Our refinements produced accurate values of the lattice parameters, even if atomic positions and thermal parameters are less reliable. From the lattice parameters at 300 and 84 K, we determined that the linear thermal expansion coefficient is $-31.1 \times 10^{-6} K^{-1}$.

To gain more insight in the mechanism behind the NTE behavior in this compound, we studied the neutron vibrational spectrum. The modes of vibration of an octahedral molecule, $X(YZ)_6$, are easily obtained. The 33 modes of vibration of $X(YZ)_6$ where Y and Z are on 4-fold axes are:

$$\Gamma = 2A_{1g} + 2E_g + 1F_{1g} + 4F_{1u} + 2F_{2g} + 2F_{2u}$$

There are now 13 distinct types of modes: 4 stretches and 9 deformations.

The INS results are shown in figures 2 and 3. The INS spectra were interpreted with the help of simple cluster calculations. Figure 1(b) shows the cluster model used for our calculation. The linkages in this compound comprise 2 atoms and have internal dynamics, which makes the situation more complex than in silicates. Therefore progress requires a reasonably complete interpretation of the vibrational spectrum of the PB analog. A structurally optimized cluster using the HyperChem package [3] was used to obtain harmonic frequencies. These results were then used with a-climax [4] to calculate the neutron vibrational spectrum of the material. Spectral interpretation was done from model compounds, group theory, group frequencies, and comparison with reference IR and Raman spectra.

Figure 2 shows the INS spectrum for $Zn_3[Fe(CN)_6]_2 \cdot xD_2O$ below 700 cm$^{-1}$. Analysis of the INS data shows well-defined, well-separated bands corresponding to the stretching and deformation modes of the Fe and Zn octahedra, all below 800 cm$^{-1}$.

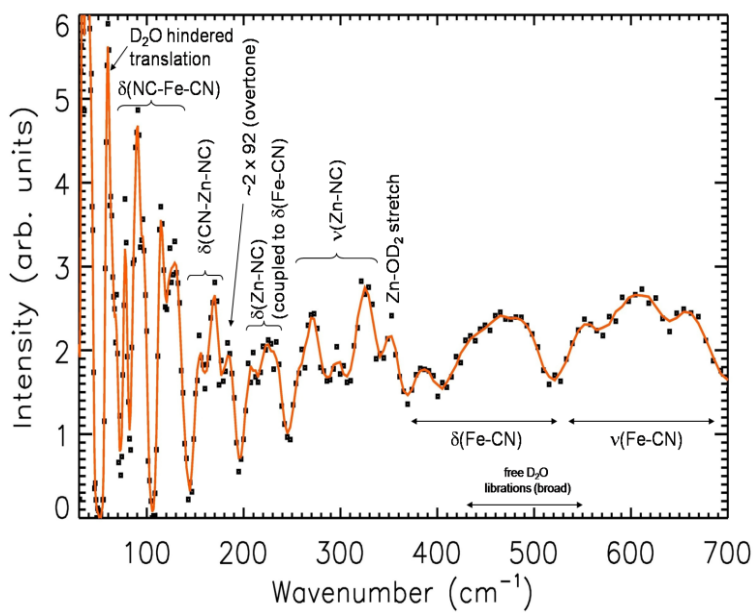

**Figure 2.** The inelastic neutron scattering spectrum (INS) of $Zn_3[Fe(CN)_6]_2 \cdot xD_2O$ below 700 cm$^{-1}$ showing the different bands of stretching and deformation modes of Fe and Zn octahedra.

The INS spectrum in the 500-2500 cm$^{-1}$ range is shown in figure 3. We observed weak CN stretches, which is not unexpected given the small scattering cross section of C and N and the weak amplitude of their motion. In $Zn_3[Fe(CN)_6]_2 \cdot 14H_2O$ [6], Zn occupies the Wyckoff position 4b with unit occupancy. By contrast, Fe, C, and N occupy Wyckoff sites 4a, 24e, and 24e, respectively –all with occupancy 2/3. Fe, C, or N sites may be occupied by oxygen (from water). If so, $Fe(CN)_6$ octahedra should be intact, but isocyanide ligands may be missing around Zn. Instead, Zn-O(H) may be present.

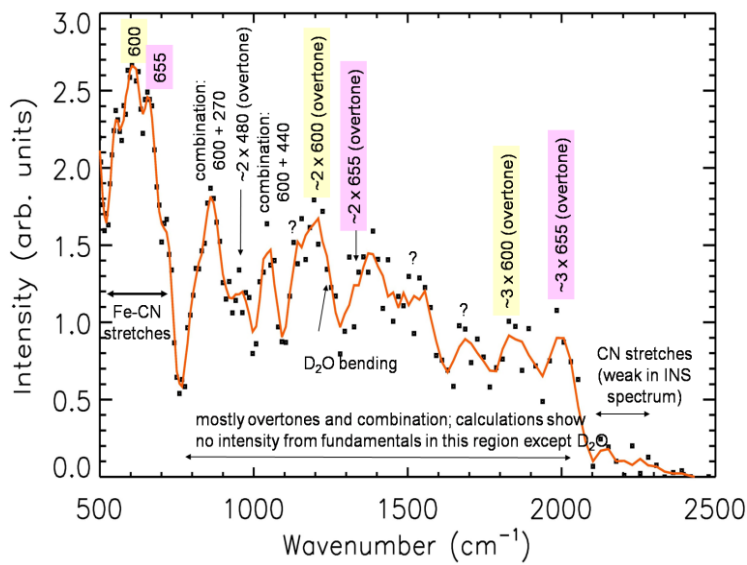

**Figure 3.** The inelastic neutron scattering spectrum (INS) of $Zn_3[Fe(CN)_6]_2 \cdot xD_2O$ within 500-2500 cm$^{-1}$ energy range showing combination, overtones and other vibrational modes.

In addition to the water at well-defined sites in the lattice, lattice water may be present in the cavities between the octahedrons. Relatively strong Zn-OD$_2$ stretching mode at 380 cm$^{-1}$ in the INS spectrum confirms that some water occupies interstitial positions in the lattice.

## 4. Conclusion

To our knowledge this is the first time the INS spectrum of a PB analog has been calculated and interpreted in greater detail. More detailed calculations of the vibrational modes are in progress to finalize the interpretation of the vibrational spectrum and help with future measurements of mode Grüneisen parameters. Indeed, this will open up the possibility of correlating NTE behavior with the temperature dependence of the mode Grüneisen parameter.

**Acknowledgement**

The Basic Energy Sciences program of the US Department of Energy supported the work at Los Alamos Neutron Science Center (LANSCE).